# Systematic control of carrier concentration and mobility in RF sputtered ZnO/Al:ZnO thin films

Navid M. S. Jahed[1,2] and Siva Sivoththaman[1,2]

[1]Centre for Advanced Photovoltaic Devices and Systems, University of Waterloo, Ontario, Canada, N2L 3G1

[2]Waterloo Institute for Nanotechnology, University of Waterloo, Ontario, Canada, N2L 3G1

**Abstract:**

RF sputtered ZnO and Al:ZnO films are attractive transparent conductive oxides for fabrication of opto-electronic devices. In this paper we present efforts to control carrier concentration and mobility of ZnO/Al:ZnO thin films by controlling deposition parameters (RF power, pressure and substrate temperature). Al:ZnO thin film with resistivity as low as $= 3.8 \times 10^{-4}$ .cm at deposition temperature of $250^0$C has been achieved. Zinc oxide thin film with low resistivity of $= 3.7 \times 10^{-2}$ .cm and high electron mobility of 30 cm$^{-2}$V$^{-1}$s$^{-1}$ at deposition temperature of $250^0$ C with acceptable electronic parameters stability has been obtained. Light transmission of Al:ZnO and ZnO samples deposited on glass at different substrate temperature has been studied. Investigation were made to assess the effect of deposition temperature on the photoluminescence spectra (PL) of ZnO/Al:ZnO sputtered on silicon and glass substrate. The evolution of near band edge (NBE) and deep level emission (DLE) photoluminescence peaks with deposition temperature in ZnO/Al:ZnO sputtered on Silicon and glass substrate have been studied.

## I Introduction:

Wide band gap transparent conductive oxides (TCOs) offer high transparency in the visible range, owing to a large intrinsic band gap (~3 eV), combined with a tunable conductivity. Conductivity in TCOs is adjustable over a wide range, from nearly insulating to highly conductive, mainly due to defect (non-stoichiometry) or dopant controlled n-type carrier concentration [1]. Common TCOs are based on tin oxide ($SnO_2$), indium oxide ($In_2O_3$), indium tin oxide (ITO), fluorine doped tin oxide (FTO) and zinc oxide (ZnO). Now it has been over a decade that TCOs based on Zinc oxide has gained attention to replace the commonly used ITO in optoelectronic devices. The driving force behind this interest comes from the fact that Zinc is a more abundant material in nature and it also can be easily produced with high purity (99.99%). As a result zinc oxide offers a cheaper (More than 10 times), nontoxic, bio compatible, chemically stable replacement for ITO. Moreover, ZnO films are less absorber in UV range and can be doped by more available group III elements (boron, aluminum, gallium or indium).

ZnO and impurity doped ZnO are finding applications in a wide range of optoelectronic devices such as: solar cells, LEDs, front panel displays, TFTs, etc. TCOs based on ZnO films can be prepared in crystalline, polycrystalline and amorphous forms. ZnO and Aluminium doped zinc oxide (Al:ZnO) thin films are prepared by a variety of techniques such as: evaporation, pulsed laser deposition, chemical deposition, Metal Organic Chemical Vapor Deposition (MOCVD), spray pyrolysis and magnetron sputtering. Among these techniques magnetron sputtering is of particular interest, since it is offering low temperature deposition, robust process, repeatability, uniformity and large area fabrication adaptability.

Sputtered Aluminium doped zinc oxide (Al:ZnO or AZnO) is often used as a transparent front contact and back reflector for light trapping in first generation solar cell technology [2]. Al:ZnO is also widely used as a transparent electrode in second generation thin film solar cells [3]. Sputtered Al:ZnO is a degenerated, semi metallic and highly conductive TCO. Resistivity of sputtered Al:ZnO has been lowered down to that of ITO ($\rho \sim 10^{-4}$ $\Omega$.cm) over the past decade. The resistivity of impurity doped zinc oxide TCOs have been reduced to the range of $10^{-5}$ $\Omega$.cm under controlled laboratory conditions by increasing the carrier concentration up to $1.5 \times 10^{21}$ cm$^{-3}$ [4, 5]. Although high carrier concentration might be desirable in terms of better electrical properties, however, in contrast it is unfavorable in terms of optical properties particularly in higher wave lengths range. Increased number of free carriers, increases light reflection due to free electrons oscillations above a wavelength known as plasma frequency. This effect is mainly prominent in near infrared & infrared regions and cause decrease in light transmission in near infrared and infrared regions. Consequently, in thin film solar cell applications or some third generation solar cell applications where a high transmission in the range of 400 nm-1300 nm is required, a control over carrier concentration and mobility to optimize reflection from free electrons oscillations above plasma frequency is desirable.

Analogous to sputtered Al:ZnO, ZnO film deposited by sputtering is a polycrystalline thin film. In contrast to Al:ZnO which possess a metallic like behavior, ZnO is more like a semiconductor with defect/stoichiometry controlled carrier concentration ranging from $10^{16}$ cm$^{-3}$ to $10^{20}$ cm$^{-3}$. Sputtered ZnO thin films are used in TFTs, LEDs and in third generation solar cells such as quantum dot sensitized solar cells (QDSSCs) [6].

In the device structures that ZnO acts as an active layer, it is important to be able to control mobility and carrier concentration in ZnO layer. For example in third generation heterojunction quantum dot sensitised solar cells (HQDSSCs), where active layer is the junction between a quantum dot layer and ZnO layer, control over carrier concentration in ZnO layer is of crucial importance [7]. In HQDSSCs light absorption happens in QD layer where usually carrier mobility is low outside of depletion region in QD layer. To have a fully depleted device, which offers better carrier mobility in depleted region and better exciton dissociation, control over carrier concentration in ZnO layer is necessary [8].

This study is an effort to control carrier concentration and hall mobility in sputtered ZnO/Al:ZnO thin films by controlling deposition conditions (power, pressure, temperature). Light transmission of the thin films has also been studied to allow optimization between electrical and optical properties in advanced solar cell applications. Effect of deposition conditions on Photoluminescence of the films has also been studied for quantum efficiency optimization in devices where ZnO/Al:ZnO is used as a light emitting active layer (such as blue LEDs, light emitting TFTs, etc.).

## II Experimental details:

Zinc oxide and Al doped zinc oxide films were deposited by radio frequency magnetron sputtering on corning glass (Eagle XG) and (100) p-type silicon wafers in Intel Vac. sputtering tool. Ceramic discs of

Al$_2$O$_3$: ZnO (2 wt %) with purity of 99.999% and diameter of 3 inch from Angstrom sciences has been used as Al:ZnO target. Similar ceramic discs of ZnO with purity of 99.999% has been used as ZnO target and installed on another gun inside the chamber. Our sputtering tool has the ability to switch power between the guns which allows target selection. Standard RCA cleaning procedures was used to clean the samples prior to loading them into the chamber. The samples are placed parallel to the targets in four sample holders at a distance of 50 ± 3mm. To improve the uniformity (> 94%) a mechanical system is designed to provide both revolution (for 4 sample holders) and rotation (for each individual sample holder) motion to the sample holders batch [9]. The sample holder batch has 40 revolution per minute while each individual sample holder possesses 14 rotation per minute. The chamber left over night to reach background pressure of $4\times10^{-7}$ torr before each deposition. The non- reactive deposition is carried out in Ar as inert gas inside the chamber. The deposition pressure was varied between 0.5 mtorr and 10 mtorr. The pressure was stabilized during the deposition by controlling the gas flow rate with mass flow controllers. The deposition power was varied in the range of 80 W to 300 W. The deposition temperature is controlled by a thermocouple placed in the vicinity of sample holders. Temperature controller is recalibrated to match the temperature readings of reference temperature logger temporarily placed on the back of the substrates for more accurate readings. The temperature is stabilized inside the chamber by 1 hour preheating the chamber at deposition temperature. Prior to every deposition, a two minutes of pre deposition (with closed shutter) is performed at 100 W to clean up the surface of the targets, followed by a 1 minute warm up at deposition power. The deposition is performed with a stabilized plasma up to 120 minutes.

The sheet resistance of the samples are measured with a four point probe system right after unloading them from the chamber. The samples are patterned by a UV lithography mask aligner system (OAI: optical associates incorporation) and wet etched in a solution of 10% HCL in water. The thickness of the samples were measured using a stylus-type profilometer (Dektak 150). The Transmission of light on samples prepared on corning glass was measured using PerkinElmer UV/VIS/NIR Lamda 1050 spectrometer with air as the reference. The open source PUMA software was used to verify the thickness on ZnO coated glasses and attain index of refraction using the transmission spectra of the samples [10]. Edinburgh Instrument Fluorescence Spectrometer was used to acquire Photoluminescence (PL) spectra at room temperature upon excitation at 310 nm from a 900 W Xe lamp. A long pass optical filter with lower cut off wave length at 340nm is placed before monochromator in emission arm to prevent formation of excitation harmonics. The carrier concentration and mobility of the films were characterized by Hall Effect experiment in a Van der pauw configuration with Ecopia HMs-300 setup in a magnetic field of 0.54 T at room temperature. To avoid temperature effect on the films instead of making contact with soldering of In-Sn or E-beam deposition of Al/Ti , silver paste is used to make an Ohmic contact on the ZnO/AZnO coated glass samples when possible. Samples were diced from the centre of 4 inch coated glass wafers. Compared to alternative ways of making Ohmic contact, silver paste was providing faster and acceptably lower resistivity Ohmic contact. Prior to every Hall measurement experiment, I-V characteristic of our samples prepared in Van der pauw configuration, is studied to make sure that the contacts are Ohmic. The Hall measurement current is adjusted accordingly for every sample to have a meaningful Hall measurement experiment.

## III Results and discussion:

*Aluminium doped zinc oxide*

Al:ZnO films can be sputtered from $Al_2O_3$:ZnO targets with different $Al_2O_3$/ZnO weight ratios, often referred as target doping concentration (TDC). Ceramic $Al_2O_3$: ZnO targets with TDCs of 0.5%, 1%, 2% and 4% are available. $Al_2O_3$:ZnO targets with 2% TDC are of particular interest since they result in the least thickness dependent resistivity in the thin films [11]. Moreover, lowest reported resistivities are achieved by targets with TDC of 2% [12].

Deposition pressure is the sputtering parameter that strongly effects electrical properties of the AZnO thin film. As a rule of thumb, at higher pressure, deposition rate is lower because of smaller mean free path in particles. In dynamic sputtering systems (with rotation and/or revolution motions) the deposition rate is several times lower than stationary systems. As a result in dynamic sputtering systems it is desirable to optimise sputtering parameters for high quality, conductive AZnO thin films at lower pressure to increase target life and reduce material waste.

Table.1 summarizes the electrical properties of AZnO samples prepared from $Al_2O_3$: ZnO targets with 2% TDC at different pressures and powers. As it is apparent in the table, deposition rate decreases by increasing pressure when the power is fixed. Sheet resistance of the samples are measured right after unloading them from the chamber. Films deposited at higher pressures show instability in electrical properties over time when left outside of the chamber at room temperature. Particularly, degradation in sheet resistance was more apparent for the films deposited at 10 mtorr and 150 W, 200 W at room temperature. This effect is reflected in Hall measurement results, which is done up to one week after samples were unloaded from the chamber and kept at room temperature inside petti dish at clean room, and is more prominent for the samples that are prepared at higher pressures and higher powers. As reported by Minami et al. this effect is mainly due to diffusion of atmospheric oxygen in the AZnO/ZnO thin film and is prominent in thinner films and humid ambient [13]. Atmospheric oxygen diffusion causes decrease in carrier concentration along with decrease in mobility which both are responsible for electrically degradation of films and increase in resistivity [14]. For Al:ZnO films deposited at pressures below 5 mtorr, this effect becomes less significant. The Films deposited at low pressure of 0.5 mtorr show very stable electrical properties when left outside of chamber and as can be seen from the table.1, measured sheet resistance right after deposition $R_{sh}$ is very close to that calculated from hall measurement results ($\rho = R_{sh} \times t$ ).

To check the effect of the deposition temperature on the stability of films deposited at high pressure, we repeated our least stable condition (150 W and 10 mtorr) at $250^0$ C. The sheet resistance of this film ($R_{sh}$ = 58 Ω/sq at 180 nm) was more stable over time. The instability of electrical properties for the films deposited at higher pressure is relaxed when films are sputtered at deposition temperature of $250^0$ C (not shown in the table) .In AZnO films carrier concentration comes from impurity dopants (Al) and non-stoichiometry (mainly in the form of oxygen vacancy). As a result diffusion of oxygen decreases number of free carriers from oxygen vacancies. It seems that for the films deposited at high pressure, aluminium dopants are not activated and oxygen vacancy is the main source of free carrier concentration. At elevated deposition temperature ($250^0$ C), aluminium dopants are activated and as a result oxygen diffusion has less effect on electrical stability of the film.

At higher pressure when we are increasing the power above 150 W, there is arcing in the plasma which causes re-deposition of AZnO from coated films on the sample holder. This re-deposition makes black spots on the samples. This effect is eliminated when deposition pressure is lowered to 0.5 mtorr. At low pressure of 0.5 mtorr and proper power of 150 W, resistivity as low as = $1.5 \times 10^{-3}$ .cm is achieved. In order to decrease the resistivity further, deposition temperature has been increased up to $250^0$ C.

Table 1 : Variation of Electronic transport parameters Sheet resistance ($R_{sh}$), Resistivity ($\rho$), Carrier concentration ($N_e$) and Hall mobility ($\mu$) of the Al:ZnO films sputtered on glass substrate at room temperature with different deposition power and pressure.

| | RF Power (Watt) | Rate of deposition ($A^0$/min) | $R_{Sh}$* ($\Omega$/sq) | Film resistivity ($\Omega$.cm) | Carrier concentration $N_e$ (cm$^{-3}$) | Hall mobility $\mu$ (cm$^{-2}$V$^{-1}$s$^{-1}$) | Time (min.) | Thickness (nm) |
|---|---|---|---|---|---|---|---|---|
| Pressure: 10 mtorr Ar flow=165 sccm | 200 | 36.67 | 504 | 6.18E-03 | 1.34E+20 | 7.563 | 60 | 220 |
| | 150 | 21.7 | 79 | 2.24E-01 | 1.79E+20 | 1.55E-01 | 60 | 130 |
| | 120 | 6.61 | 1505 | 2.95E-01 | 4.829E+20 | 4.38E-02 | 120 | 80 |
| Pressure: 5 mtorr Ar flow=120 sccm | 240 | 70 | 130 | 2.24E-02 | 6.19E+19 | 4.496 | 60 | 420 |
| | 150 | 41.7 | 249 | 4.70E-02 | 6.60E+19 | 2.018 | 60 | 250 |
| | 80 | 18.3 | 287 | 3.09E-03 | 9.30E+19 | 2.17E-01 | 60 | 110 |
| Pressure: 2 mtorr Ar flow=80 sccm | 300 | 125 | 51.6 | 3.33E-02 | 1.817E+20 | 1.032 | 60 | 750 |
| | 200 | 49.3 | 18.8 | 1.33E-01 | 1.19E+19 | 3.956 | 150 | 740 |
| | 100 | 36.7 | 263 | 3.62E-03 | 1.75E+20 | 9.834 | 60 | 220 |
| Pressure: 0.5 mtorr Ar flow=50 sccm | 200 | 79.2 | 25.2 | 2.017E-03 | 3.75E+20 | 8.182 | 120 | 950 |
| | 150 | 56.7 | 27.2 | 1.52E-03 | 5.51E+20 | 7.57 | 120 | 680 |
| | 100 | 36.25 | 43.4 | 1.9E-03 | 2.67E+20 | 8.805 | 120 | 435 |

* Sheet resistance of all of the samples are measured right after deposition whereas hall experiment is performed after a week.

The effect of deposition temperature on the electronic transport parameters of AZnO thin films is summarized in table2. As can be seen from the table, the steady decrease in resistivity by increasing temperature is because of increasing trend in both mobility and carrier concentration with respect to temperature. AZnO films with resistivity as low as $\rho = 3.8 \times 10^{-4}$ $\Omega$.cm has been achieved. The optical transmissions of the films are represented in fig.1. Transmission of light in visible range is above 85% for all of the films. As it can be seen from the fig.1, by increasing the deposition temperature we have a continuous decrease in light transmission in the infrared region. This decrease in light transmission in infrared range is mainly because of increasing trend that is observed in carrier concentration by increasing deposition temperature.

One important signature of every material is its near band edge photoluminescence (NBE). There are several mechanisms responsible for the occurrence of near band edge peaks, like free exciton emission (FX), bound exciton (BX), donor-acceptor pair (DAP), etc. All of these mechanisms have excitonic behavior and are temperature dependent. Zinc oxide has a large exciton binding energy (~ 60 meV) which makes it possible to have free exciton emission up to room temperature [15]. An important precondition to have NBE emission at room temperature is the suppression of non-radiative processes which corresponds with better crystalline quality. As a result, presence of NBE emission is an indication of better crystalline quality. In addition to NBE emission, zinc oxide has also deep level emissions (DLE) in the range of 405 nm to 750 nm. The origin of broad band DLEs is still under debate but it is believed that defects in the form of zinc vacancy/interstitial and oxygen vacancy/interstitial are the main cause for these so called defect related emissions.

Table 2 : Variation of electronic transport parameters Sheet resistance (Rsh), Resistivity ( ), Carrier concentration (Ne) and Hall mobility (µ) and etching rate (10% HCl in water) of AZnO films sputtered at 0.5 mtorr and 150 W with different deposition temperature.

| Temperature ($C^0$) | $R_{Sh}$ ( /sq) | Film resistivity ( .cm) | Carrier concentration $N_e$ ($cm^{-3}$) | Hall mobility µ ($cm^{-2}V^{-1}s^{-1}$) | Thickness (nm) | Index of refraction (n) at 550nm | Etching rate in 10%HCL ($A^0/S$) |
|---|---|---|---|---|---|---|---|
| 250 | 5.8 | 3.8E-04 | 1.07E+21 | 15 | 780 | 1.86 | 11.1 |
| 200 | 8.3 | 6.13E-04 | 8.1E+20 | 12.7 | 775 | 1.85 | 11.4 |
| 180 | 8.6 | 6.67E-04 | 5.57E+20 | 16.4 | 775 | 1.85 | 12.9 |
| 160 | 11.8 | 7.21E-04 | 8.17E+20 | 11 | 770 | 1.85 | 14.0 |
| 140 | 12.5 | 9.23E-04 | 5.42E+20 | 12.5 | 765 | 1.84 | 15.3 |
| 120 | 16.6 | 1.14E-03 | 5.91E+20 | 9.46 | 760 | 1.89 | 16.9 |
| 90 | 23 | 1.51E-03 | 6.68E+20 | 6.35 | 725 | 1.86 | 20.7 |
| 60 | 25.8 | 1.71E-03 | 6.71E+20 | 5.43 | 710 | 1.86 | 23.7 |
| RT | 27.2 | 1.52E-03 | 5.51E+20 | 7.57 | 680 | 1.85 | 29.0 |

Room temperature photo luminescence spectra of Al:ZnO films deposited at 150 W and 0.5 mtorr at different deposition temperatures on silicon and glass substrates are shown in fig.2 and fig.3 respectively. Sputtered Al:ZnO on Si substrate deposited at room temperature (pale blue line in fig.2) exhibits NBE peaks at 367 nm and 396 nm with DLE peaks at 426 nm,468 nm (blue), 653 nm and 740 nm (red). The peak at 620 nm is the artifact of the measurement system that comes from second harmonic of the excitation source (at 310 nm) and is not photoluminescence emission of the samples. As can be seen from the fig.2, by increasing the deposition temperature, the position of NBE and DLE peaks gradually changes. At deposition temperature of $250^0$ C, NBE emission peaks are located at 363 nm and 390 nm whereas DLE peaks are located at 426 nm, 485 nm (blue spectrum range), 595 nm (orange spectrum rage), 678 nm and 731 nm (red spectrum range). Ahn et al., using full potential linear muffin-tin orbital method has calculated the defects' energy levels responsible for DLEs in zinc oxide materials [16].

According to Ahn et al., DLE peak at 426 nm can be attributed to zinc vacancies, the peak at 595 nm and 615 nm can be attributed to oxygen interstitials. Interesting, is the joint behavior of DLE pairs at 653 nm, 740 nm for the sample prepared at room temperature and DLE pairs at 678 nm, 731 nm for the sample prepared at $250^0$ C. As shown by Alvi et al., using full potential linear muffin-tin orbital method, these peaks can be attributed to oxygen vacancies situated ~ 1.65 eV below the conduction band [17]. DLE peaks at 468 nm and 485 nm can be attributed to negatively charged zinc interstitials situated ~2.5-2.6 eV below the conduction band [18]. These changes in the type of defects by increasing the deposition temperature is an indication of tremendous structural change in the thin film of Al:ZnO. One noticeable fact about fig.2 is the steady increase, neglecting drops at $90^0$ C, in the photoluminescence emission of samples by increasing deposition temperature. Of particular interest is the ratio of highest NBE emission to highest DLE ($I_{NBE}/I_{DLE}$) This ratio has steadily been increased from 1.7 for room temperature deposition to 3 for deposition temperature of $250^0$ C (maximum of $I_{DLE}$ at 485 nm). This means that crystalline quality of the sample has been improved by increasing deposition temperature.

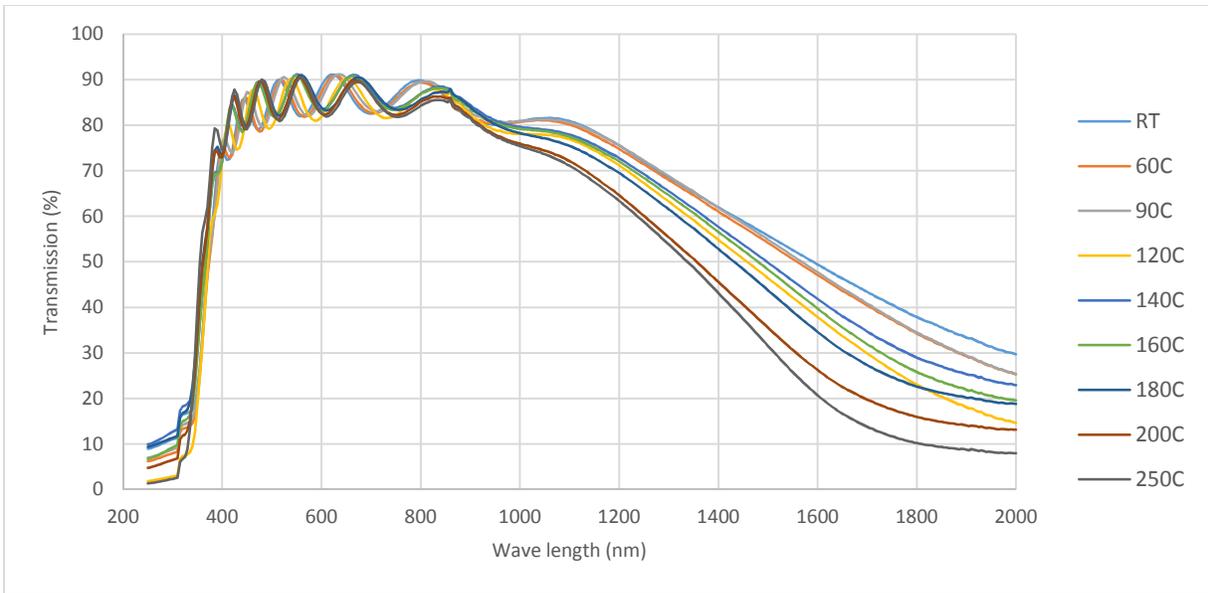

Figure 1: Change in Transmission of Al:ZnO film deposited at pressure = 0.5 mtorr, Power = 150 W with respect to deposition temperature.

The effect of deposition temperature on photoluminescence of Al:ZnO samples prepared on glass substrate is even more drastic. As depicted in fig.3, the photoluminescence intensity increases by two orders of magnitude for sample prepared at substrate temperature of $160^\circ$ C. A congruous behavior is observed in electronic transport parameters (Table 2) where at substrate temperature of $160^0$ C there is an apex in carrier concentration and a drop in hall mobility.

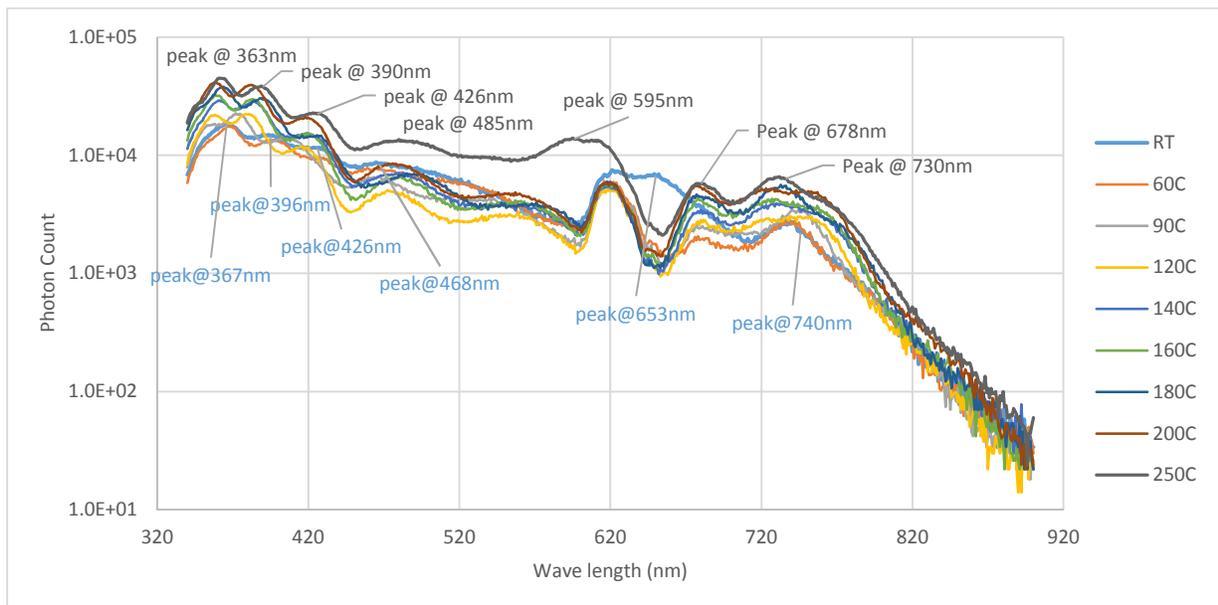

Figure 2: Room temperature photo luminescence spectra evolution of Al:ZnO film deposited at 150 W and 0.5 mtorr on Si substrate with different deposition temperatures in logarithmic scale. Excitation wave length is 310 nm. The emission peaks for sample prepared at substrate temperature of $250^0$ C is depicted.

The change in position of emission peaks along with change in photoluminescence intensity is an indication that samples have gone through a tremendous structural change at deposition temperatures between $140^0$ C-$160^0$ C [19]. The $I_{NBE}/I_{DLE}$ ratio for samples prepared on glass substrate at RT, $160^0$ C and $250^0$ C is 0.003, 0.762 and 2.95 respectively which is an indication of an improvement in crystalline quality. All of the samples, regardless of the substrate being glass or silicon, show a strong near band edge photoluminescence emission (NBE) at deposition temperatures above $160^0$ C.

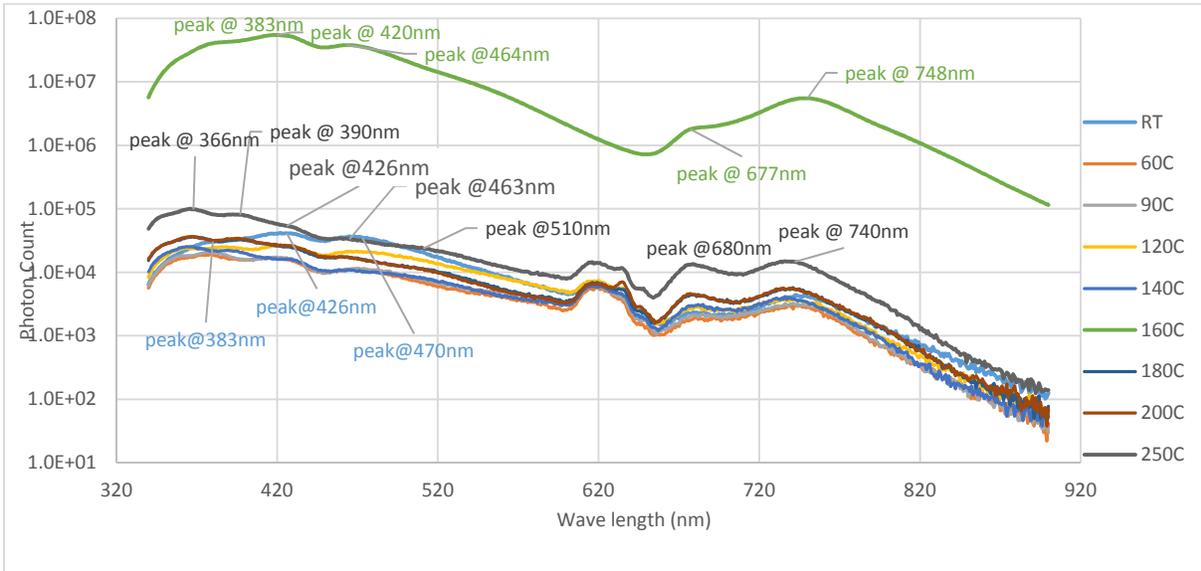

Figure 3: Room temperature photo luminescence spectra evolution of Al:ZnO films deposited at 150 W and 0.5 mtorr on glass substrate with different deposition temperatures in logarithmic scale. Excitation wave length is 310 nm. The emission peaks for samples prepared at substrate temperature of $160^0$ C and $250^0$ C is depicted.

*Zinc oxide*

Table 3 summarizes the electrical properties of zinc oxide samples prepared from ZnO target at different chamber pressure and deposition power. Like sputtered aluminium doped zinc oxide, sputtered ZnO films represent sensitivity to atmospheric oxygen when exposed to air. In comparison to Al:ZnO samples, the degradation in electrical properties due to atmospheric oxygen diffusion is more significant in ZnO samples where even ZnO films deposited at low pressure show degradation in electrical properties right after being exposed to air.

The effect of deposition temperature on electronic transport parameters of ZnO films sputtered at 150 W and 0.5 mtorr is summarized in Table.4. Deposition temperature has a drastic effect on electrical properties of ZnO films which is reflected in the decrease of sheet resistance of the films (measured right after unloading the samples from chamber) by increasing substrate temperature. Except for the film deposited at substrate temperature of $250^0$ C and $200^0$ C, the rest of the ZnO films degrade considerably when exposed to air. By increasing substrate temperature up to $250^0$ C, ZnO films with resistivity as low as $= 3.7 \times 10^{-2}$ .cm with acceptable stability in air ambient and room temperature is achieved. As depicted in fig.4 the higher mobility and lower carrier concentration of these films allows for better light transmission in the infrared range. All of the ZnO samples show transmission higher that 85% in the visible range. The change in the carrier concentration is consistent with change in transmission spectra in the infrared range and is due to change in both carrier concentration and mobility however, in comparison to Al:ZnO, the changes in mobility is more significant than that in Al:ZnO.

Table 3: Variation of Electronic transport parameters Sheet resistance (Rsh), Resistivity ( ), Carrier concentration (Ne) and Hall mobility (μ) of the ZnO films sputtered on glass substrate at room temperature with different deposition power and pressure.

| | RF Power (Watt) | Rate of deposition ($A^0$/min) | $R_{Sh}$* ( /sq) | Film resistivity ( .cm) | Carrier concentration $N_e$ (cm$^{-3}$) | Hall mobility μ (cm$^{-2}$V$^{-1}$s$^{-1}$) | Time (min.) | Thickness (nm) |
|---|---|---|---|---|---|---|---|---|
| Pressure: 10 mtorr Ar flow=165 sccm | 200 | 35.8 | 510 | 2.6E+02 | 1.11E+16 | 2.31 | 120 | 430 |
| | 150 | 17.9 | 412.6 | 4.51 | 3.16E+16 | 41 | 120 | 215 |
| | 120 | 6.25 | 4310 | 5.1 | 1.45E+18 | 8.4E-1 | 120 | 75 |
| Pressure: 5 mtorr Ar flow=120 sccm | 200 | 57 | 9850 | 27 | 2.28E+17 | 1.1 | 60 | 340 |
| | 150 | 43 | 284 | 1.6E+02 | 6.70E+16 | 5.52E-01 | 75 | 325 |
| | 100 | 21 | 12800 | 4.3E+03 | 1.30E+16 | 1.12E-01 | 60 | 126 |
| Pressure: 2 mtorr Ar flow=80 sccm | 200 | 125 | 61300 | 1.58E+03 | 3.55E+16 | 1.1E-01 | 120 | 620 |
| | 150 | 39.5 | 2900 | 2.10E-01 | 1.70E+17 | 16.71 | 120 | 475 |
| | 100 | 36.1 | 1.3E+5 | 74 | 5.9E+17 | 14.30 | 65 | 235 |
| Pressure: 0.5 mtorr Ar flow=50 sccm | 200 | 75 | 180.3 | 1.5 | 2.3E+18 | 1.41 | 90 | 680 |
| | 150 | 60 | 90000 | 11.8 | 2.1E+17 | 2.2 | 120 | 720 |
| | 100 | 35 | 144 | 7.00E-01 † | 9.5E+17 † | 9.4 † | 100 | 360 |

* Sheet resistance of all of the samples are measured right after deposition whereas hall experiment is performed after a week.
† Hall measurement for this sample is done a few hours after unloading it from chamber.

Fig.5 and fig.6 represent room temperature photoluminescence spectra of ZnO samples deposited on silicon and glass substrate respectively at different deposition temperatures. The overall PL intensity of ZnO thin films are higher than that of Al:ZnO films, for both of the films deposited on silicon and glass substrate. This was expected since generally ZnO thin films show better crystalline quality than Aluminum doped Zinc oxide thin film where defect states caused by aluminium dopants can provide non radiative recombination path to suppress the overall PL intensity [20]. The PL spectra in fig.5 and fig.6 indicate that for ZnO samples prepared on either silicon or glass substrate, a drastic structural change of material starts in rather low deposition temperature of $60^0$ C and at temperatures around $140^0$ C-$160^0$ C the overall PL intensity increases one order of magnitude. Similar to Al:ZnO deposited on silicon substrate at room temperature which shows a noticeable NBE PL intensity, ZnO thin film prepared on silicon substrate at room temperature has a considerable NBE intensity. The $I_{NBE}/I_{DLE}$ ratio has steadily increased for ZnO samples deposited on silicon substrate from 2.71 for the sample deposited at room temperature to 3.43 for the sample deposited at $250^0$ C.

Table 4: Variation of electronic transport parameters Sheet resistance (Rsh), Resistivity ( ), Carrier concentration (Ne) and Hall mobility (μ) and etching rate (10% HCl in water) of ZnO films sputtered at 0.5 mtorr and 150 W with different deposition temperature.

| Temperature ($C^0$) | $R_{Sh}$* ( /sq) | Film resistivity ( .cm) | Carrier concentration $N_e$ (cm$^{-3}$) | Hall mobility μ ( cm$^{-2}$V$^{-1}$s$^{-1}$) | Thickness (nm) | Index of refraction (n) at 550 nm | Etching rate in 10% HCL (A$^0$/S) |
|---|---|---|---|---|---|---|---|
| 250 † | 448 | 3.7E-02 | 5.62E+18 | 30 | 740 | 1.89 | 528.57 |
| 200 | 1880 | 1.90E-01 | 4.40E+18 | 13.1 | 500 | 1.87 | 552.22 |
| 180 | 2450 | 8.70E-01 | 1.01E+18 | 7.1 | 480 | 1.85 | 533.3 |
| 160 | 3100 | 4.23 | 2.41E+18 | 11 | 470 | 1.76 | 587.5 |
| 140 | 3636 | 1.56 | 1.78E+18 | 6.8 | 460 | 1.84 | 657.14 |
| 120 | 5220 | 1.39 | 8.31E+17 | 5.4 | 440 | 1.90 | 733.3 |
| 90 | 6520 | 6.2 | 1.86E+17 | 5.1 | 435 | 1.83 | 876 |
| 60 | 7610 | 6.5 | 2.00E+17 | 4.5 | 430 | 1.86 | 860 |
| RT† | 90000 | 11.8 | 2.1E+17 | 2.2 | 720 | 1.94 | 1440 |

* Sheet resistance of all of the samples are measured right after deposition whereas hall experiment is performed after a week.
† Deposition time for these film is 120 minute and for the rest of the films is 75 minute.

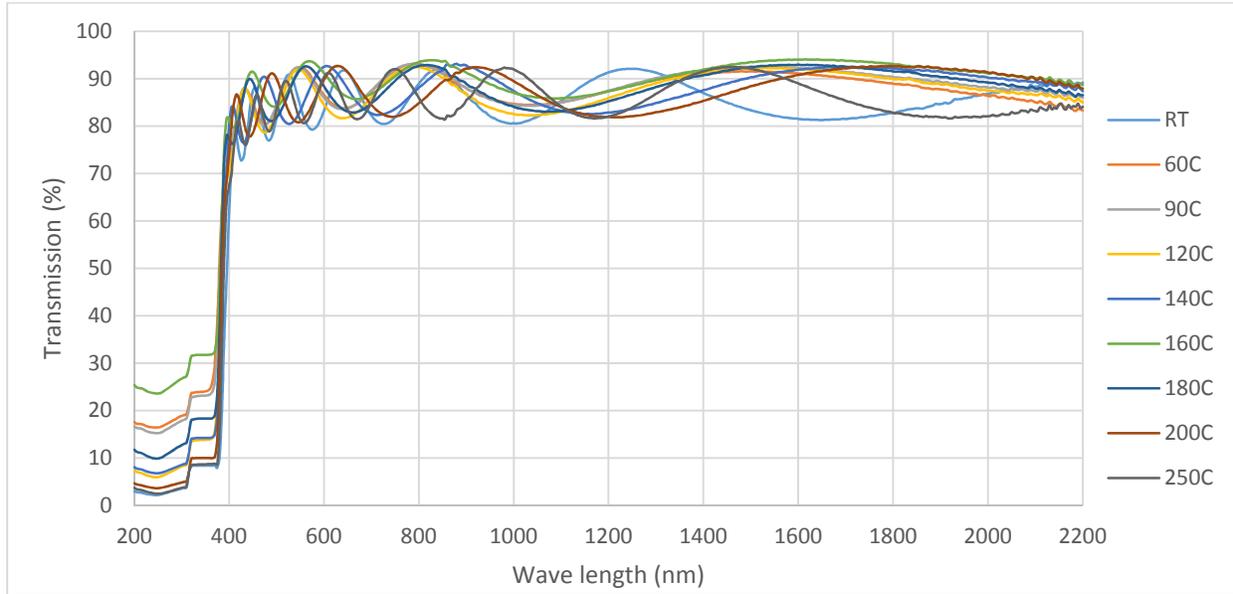

Figure 4: Change in Transmission of ZnO film deposited at pressure=0.5mtorr, Power=150 W with respect to deposition temperature. Deposition time for samples sputtered at $250^0$ C and RT is 120 min and for the rest of the films is 75 min.

In contrast to samples prepared on silicon substrate at room temperature, where NBE peaks are noticeable, both ZnO and Al:ZnO samples prepared on glass substrate show suppressed PL intensity for near band edge emissions (NBE). The $I_{NBE}/I_{DLE}$ ratio for ZnO sputtered on glass substrate has increased from 0.68 for room temperature deposition to 1.7 for deposition at $250^0$ C. like Aluminium doped ZnO thin films, all of the zinc oxide samples, regardless of type of substrate, prepared at deposition temperatures above $140^0$ C show noticeable NBEs.

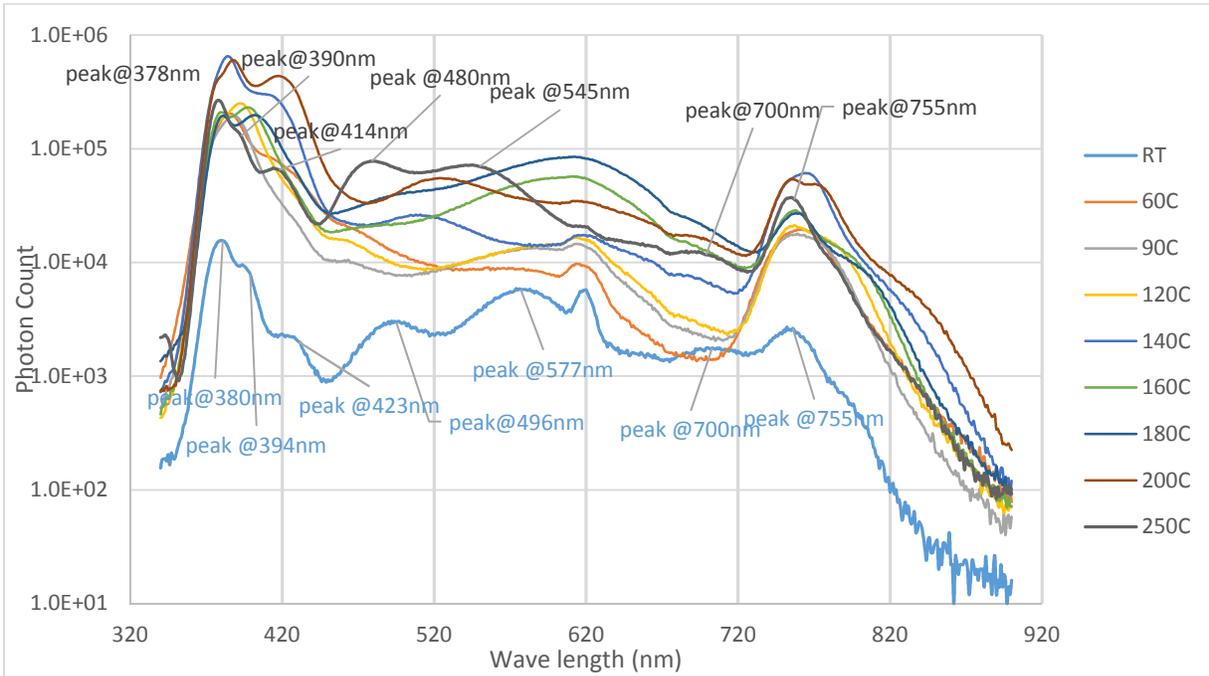

Figure 5: Room temperature photo luminescence spectra evolution of ZnO films deposited at 150 W and 0.5 mtorr on Si substrate with different deposition temperatures in logarithmic scale. Excitation wave length is 310 nm. The emission peaks for sample prepared at $250^0$ C and room temperature is depicted.

## IV Summary

In this work electrical and optical properties of zinc oxide and aluminium doped zinc oxide at different deposition power and pressure has been studied. Al:ZnO with resistivity as low as $= 1.52 \times 10^{-3}$ .cm at deposition temperature of RT with stable electrical properties in air has been achieved. By increasing deposition temperature to $250^0$ C, the resistivity of Al:ZnO films has been lowered to $= 3.8 \times 10^{-4}$ .cm which has acceptably high light transmission both in visible and near infrared ranges. The photoluminescence of Al:ZnO on silicon and glass substrate at different deposition temperatures has been studied. It has been found that Al:ZnO undergoes a tremendous structural changes that effects the NBE emission intensity and location of DLE peaks. According to photoluminescence signature of Al:ZnO, samples prepared with substrate temperatures above $140^0$ C show distinct NBE emission peaks.

The effect of deposition power and pressure on electrical and optical properties of ZnO have also been studied. Sputtered ZnO thin film with as deposited sheet resistance value of 144 Ohm/square has been achieved which shows high sensitivity to diffused atmospheric oxygen. ZnO thin films with stable electronic transport parameters only achieved at relatively high deposition temperature of $200^0$ C-$250^0$ C. The effect of substrate temperature on photoluminescence spectra of sputtered ZnO samples have been studied. Regardless of substrate type, all deposited ZnO and Al:ZnO samples, sputtered at power of 150 W and pressure of 0.5 mtorr, show noticeable NBE emission peaks at substrate temperatures above $140^0$ C.

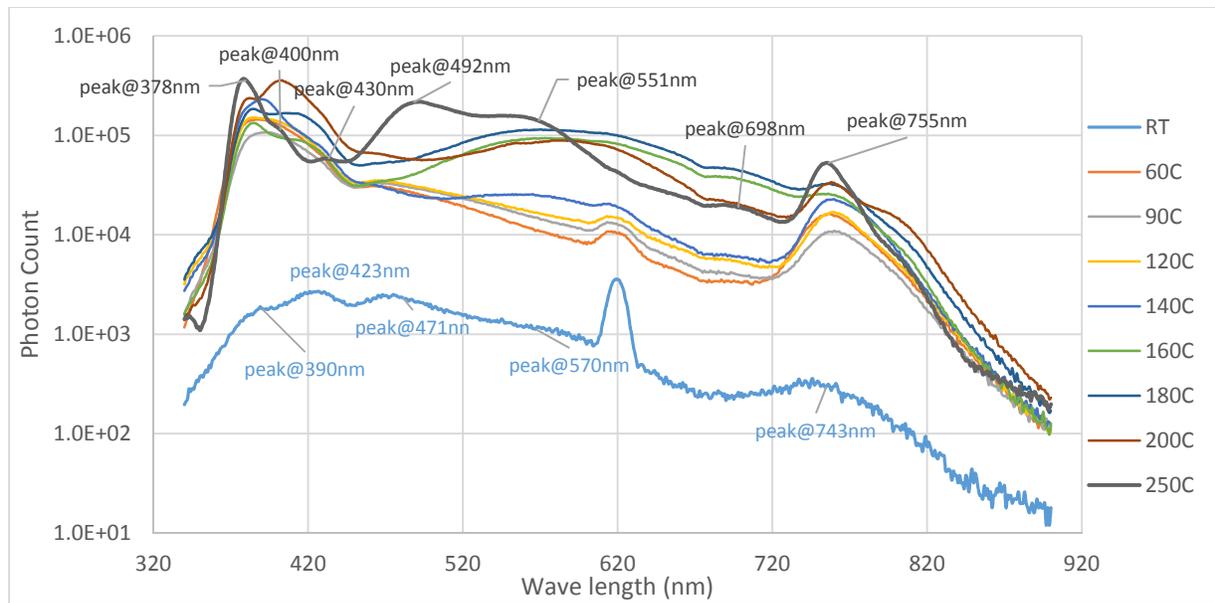

Figure 6: Room temperature photo luminescence spectra evolution of ZnO films deposited at 150 W and 0.5mtorr on glass substrate with different deposition temperatures in logarithmic scale. Excitation wave length is 310 nm. The emission peaks for samples prepared at $250^0$ C and room temperature is shown.